%% file: FlexibleAssociation_camera_ready.tex
\documentclass[conference]{./lib/IEEEtran}

\ifCLASSINFOpdf
  \usepackage[pdftex]{graphicx}
\else
\fi
\usepackage[cmex10]{amsmath}
\usepackage{amssymb}
\usepackage{mathtools}
\usepackage{array}

\usepackage{algpseudocode,algorithm,algorithmicx}
\usepackage{float}
\usepackage{tabularx}
\usepackage{mathrsfs} 
\usepackage{psfrag}
\usepackage{tikz}
\usetikzlibrary{chains, arrows,shapes,positioning,arrows,decorations.markings,calc}
\usepackage{pgfplots}
\pgfplotsset{compat=newest}
\usepgfplotslibrary{groupplots}
\usepackage{gensymb}
\usepackage{xcolor}
\usepackage[shortcuts,acronym]{glossaries}
\makeglossaries 
\usepackage{fixmath}
\usepackage{amssymb}
\usepackage{bm}
\renewcommand{\arraystretch}{2}
\usepackage{pbox}
\usepackage{algorithm}
\usepackage{lettrine}

\DeclareMathOperator*{\argmax}{arg\,max}
\renewcommand{\arraystretch}{2}	
\usepackage{url}

\algnewcommand{\IIf}[1]{\State\algorithmicif\ #1\ \algorithmicthen}
\algnewcommand{\EndIIf}{\unskip\ \algorithmicend\ \algorithmicif}

\begin{document}
%
\title{MEC-aware Cell Association for 5G Heterogeneous Networks}
\author{\IEEEauthorblockN{Mustafa Emara, Miltiades C. Filippou, Dario Sabella}
\IEEEauthorblockA{Next Generation and Standards, Intel Deutschland GmbH, Neubiberg, Germany \\
Email:$\{$mustafa.emara, miltiadis.filippou, dario.sabella$\}$@intel.com}
}
\maketitle

\input{miscellaneous/Abbreviations}
\input{introduction/introduction}

\input{system_model/system_model}
\input{flexible_cell_association/flexible_cell_association}
\input{simulation_results/simulation_results}
\input{conclusion/conclusion}
\input{literature/literature}

\end{document}

%% file: miscellaneous/Abbreviations.tex
\newacronym{CCDF}{CCDF}{Complementary Cumulative Distribution Function}
\newacronym{C-RAN}{C-RAN}{Cloud Radio Access Network}
\newacronym{HetNet}{HetNet}{Heterogeneous Network}
\newacronym{DL}{DL}{Downlink}
\newacronym{eNB}{eNB}{Evolved NodeB}
\newacronym{E-PDB}{E-PDB}{Extended-Packet Delay Budget}
\newacronym{FDMA}{FDMA}{Frequency Division Multiple Access}
\newacronym{IoT}{IoT}{Internet of Things}
\newacronym{MEC}{MEC}{Multi-access Edge Computing}
\newacronym{M2M}{M2M}{Machine-to-machine}
\newacronym{PDF}{PDF}{Probability Density Function}
\newacronym{PPP}{PPP}{Poisson Point Process}

\newacronym{QoS}{QoS}{Quality of Service}
\newacronym{RSRP}{RSRP}{Reference Signal Received Power}
\newacronym{SINR}{SINR}{Signal to Interference plus Noise Ratio}
\newacronym{TDMA}{TDMA}{Time Division Multiple Access}
\newacronym{UE}{UE}{User Equipment}
\newacronym{UL}{UL}{Uplink}

%% file: introduction/introduction.tex
\begin{abstract}

The need for efficient use of network resources is continuously increasing with the grow of traffic demand, however, current mobile systems have been planned and deployed so far with the mere aim of enhancing radio coverage and capacity. Unfortunately, this approach is not sustainable anymore, as 5G communication systems will have to cope with huge amounts of traffic, heterogeneous in terms of latency among other Quality-of-Service (QoS) requirements. Moreover, the advent of Multi-access Edge Computing (MEC) brings up the need to more efficiently plan and dimension network deployment by means of jointly exploiting the available radio and processing resources. From this standpoint, advanced cell association of users can play a key role for 5G systems. Focusing on a \ac{HetNet}, this paper proposes a comparison between state-of-the-art (i.e., radio-only) and MEC-aware cell association rules, taking the scenario of task offloading in the \ac{UL} as an example. Numerical evaluations show that the proposed cell association rule provides nearly 60\% latency reduction, as compared to its standard, radio-exclusive counterpart.

\end{abstract}
%
%
%
%
%
\section{Introduction}
\subsection{Motivation}

The evolution of mobile networks is characterized by a growing traffic demand (currently dominated by video content \cite{CWPFebruary2017}) and by a paradigm shift in the consumed services, where content sharing and social behavior are redefining network utilization. Moreover, the introduction of 5G systems in the near future will witness a dramatic increase of \ac{M2M} connections \cite{CWPJune2016}, due to the progressive introduction of \ac{IoT} traffic and services, that will be dominated by several new vertical business segments \cite{VWPFebruary2015} (e.g., automotive and mobility, factories of the future, health-care, media and entertainment, energy). As a consequence, 5G networks will need to effectively support huge amounts of traffic streams, both heterogeneous and variable in space and in time. 

At the same time, the emergence of \ac{MEC} will introduce computing capabilities at the edge of the network and will provide an open environment targeting low packet delays due to close proximity to end users \cite{ETSISeptember2015}.
\begin{figure}
	\begin{center}
		\input{introduction/figures/used_cases/used_cases.tex}
		\caption{Envisioned 5G reference system.}
		\label{fig:used_cases}
	\end{center}
\end{figure}
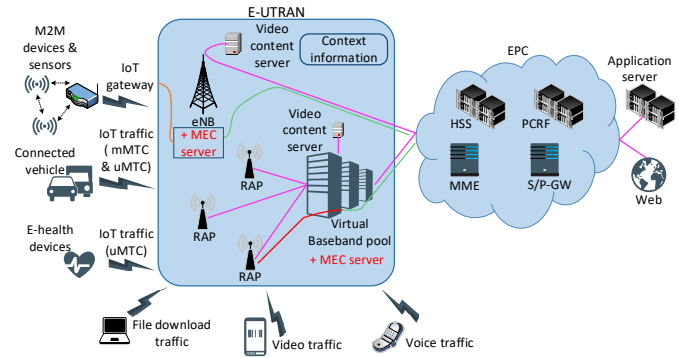
Such a challenging system scenario, as depicted in Figure \ref{fig:used_cases}, also involves a multitude of heterogeneous devices, characterized by dissimilar latency requirements, among others. The heterogeneity of QoS demands creates the need for network operators to efficiently plan and dimension the overall system by jointly capitalizing on the available communication and computation resources. 

From that perspective, the applied rule for user-cell association plays a key role towards efficiently exploiting the entire set of resources. Nevertheless, current mobile systems have been planned and deployed so far by following traditional paradigms of network planning (e.g., based on radio-only coverage). Unfortunately, this approach is not sustainable anymore, as current cell association rules completely discard the aforementioned availability of \emph{processing} resources at the network's edge, hence, they fail to constitute cost-effective and flexible solutions for QoS provisioning.

\subsection{Prior Work}
To the best of our knowledge, current technical literature mostly sheds light on the problem of optimally allocating radio and computational resources to already connected users, inherently assuming \emph{conventional} cell association, where the \ac{UE} is connected to its serving \ac{eNB} based on the maximum \ac{RSRP} rule. For example, authors in \cite{Sato2017} investigated task offloading in a multi-cell scenario, where they showed an enhancement achieved by offloading to multiple \ac{eNB}s via benefiting from prior knowledge of radio statistics. In \cite{Le2017}, the problem of radio and computational resource allocation over connected users was investigated under \ac{TDMA} and \ac{FDMA} schemes. The authors optimized the joint allocation and showed the achieved gains, as compared to a baseline round-robin scheme. Moreover, \cite{Mao2017} studied the problem of joint radio and processing power allocation under an optimization framework, where the task completion time was minimized subject to energy consumption constraints. It is, thus, evident that none of the above works questioned the effectiveness of the applied cell association rule.

With regards to the design of a cell association rule driven by performance requirements, in \cite{Li2017}, a cross-layer, ``\ac{UE} matching'' problem was studied for a \ac{C-RAN}. In this work, the authors proposed a joint matching scheme between the UEs, \ac{C-RAN} components and \ac{MEC} hosts, aiming at meeting a task completion deadline at the UE side. Nevertheless, this work did not exploit the multi-tier resource disparity expected in a \ac{HetNet} as well as reveal the practicality of the association procedure from a signaling overhead viewpoint. 

\subsection{Contribution}
Given the above described situation and identified gaps, this paper presents the following:
\begin{itemize}
	\item Focusing on a \ac{MEC}-enabled \ac{HetNet}, we introduce a new, \ac{UE}-cell association metric, which evaluates the proximity of \ac{MEC} resources to a \ac{UE}.
	\item To highlight the benefits of the proposed association rule, we introduce an \ac{E-PDB} metric, which is a one-way latency consisting of the radio transmission time of an input packet between the \ac{UE} and the connected \ac{eNB} in the \ac{UL}, along with the execution time of a given task at a \ac{MEC} host. 
	\item  We conduct numerical evaluation to compare the proposed association rule to the conventional \ac{RSRP} rule, in terms of \ac{E-PDB} performance for various inter-tier resource disparities, as well as for different network deployment densities.
\end{itemize}


The remainder of this paper is organized as follows: in Section \ref{sec:system_model}, we present an overview of the studied system model; Section \ref{sec:flexible_cell_association} elaborates on the concept of flexible cell association and Section \ref{sec:simulation_results} shows the relevant numerical results. Finally, Section \ref{sec:Conclusion} concludes the paper.

%% file: introduction/figures/used_cases/used_cases.tex
\begingroup%
  \makeatletter%
  \providecommand\color[2][]{%
    \errmessage{(Inkscape) Color is used for the text in Inkscape, but the package 'color.sty' is not loaded}%
    \renewcommand\color[2][]{}%
  }%
  \providecommand\transparent[1]{%
    \errmessage{(Inkscape) Transparency is used (non-zero) for the text in Inkscape, but the package 'transparent.sty' is not loaded}%
    \renewcommand\transparent[1]{}%
  }%
  \providecommand\rotatebox[2]{#2}%
  \ifx\svgwidth\undefined%
    \setlength{\unitlength}{260bp}%
    \ifx\svgscale\undefined%
      \relax%
    \else%
      \setlength{\unitlength}{\unitlength * \real{\svgscale}}%
    \fi%
  \else%
    \setlength{\unitlength}{\svgwidth}%
  \fi%
  \global\let\svgwidth\undefined%
  \global\let\svgscale\undefined%
  \makeatother%
  \begin{picture}(1,0.53)%
    \put(0,0){\includegraphics[width=\columnwidth]{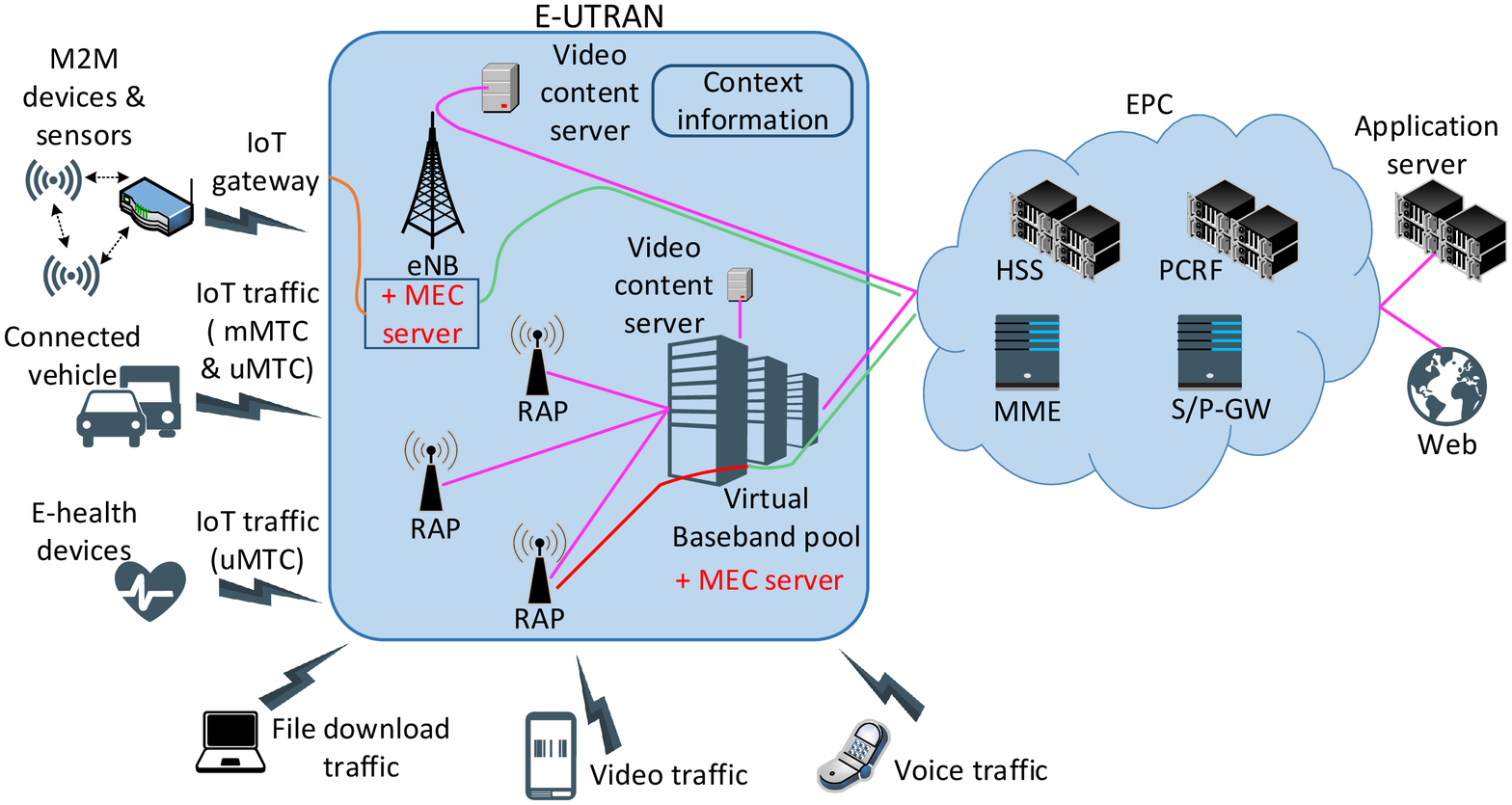}}%
  \end{picture}%
\endgroup%

%% file: system_model/system_model.tex
\section{System Model}\label{sec:system_model}

\subsection{Modeling Assumptions}

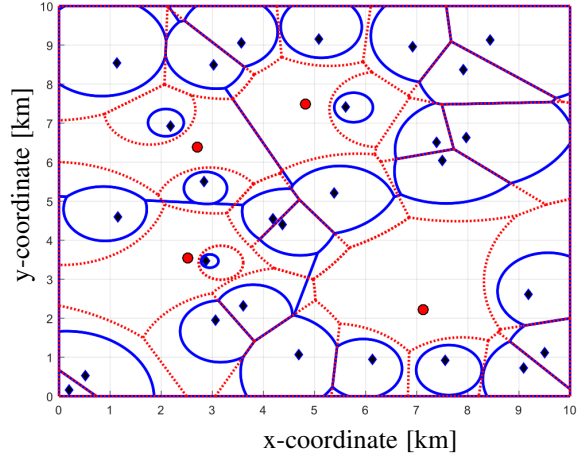
\begin{figure}
	\begin{center}
		\input{system_model/figures/system_deployment/system_deployment_2.tex}
		\caption{A two-tier network consisting of macro eNBs (red circles) of spatial density $\lambda_1 = 0.5$ eNBs/km and micro eNBs (black rhombuses) ($\lambda_2 = 6\lambda_1 , \rho_1 = 40 \; \& \; \gamma_1 = 5$). The solid blue lines represent the boundaries of the radio coverage areas determined by applying the maximum DL \ac{RSRP} association rule, while the dashed red lines represent the boundaries of the MEC coverage areas when applying the proposed, MEC-aware UE-cell association rule, which will be analyzed in Section \ref{sec:flexible_cell_association}.}
		\label{fig:system_deployment}
	\end{center}
\end{figure}

Throughout this work, a $K$-tier cellular network, such as the one illustrated in Figure \ref{fig:system_deployment}, is studied, where the \ac{eNB}s and \ac{UE}s locations are spatially randomized. According to this model, the locations of the \ac{eNB}s of the $i$-th tier are modeled through a homogeneous \ac{PPP} $\bm{\Phi_i} = \{x_i\}, i=1,2,\cdots, K $ of density $\lambda_i$, where $x_i \in \mathbb{R}^2$ represents the tier-$i$ \ac{eNB} position on a two-dimensional plane. It should be noted that the $K$ \ac{PPP}s are mutually independent. On the other hand, the \ac{UE} positions are modeled via a different, independent homogeneous \ac{PPP}, $\bm{\psi}$, of density $\lambda_u$. Due to the network's heterogeneity, different tiers are distinguished by the transmit power, $P_i$, of their \ac{eNB}s, their spatial density, $\lambda_i$, and the total processing power, $C_i$, of a \ac{MEC} server co-located with an $i$-th tier \ac{eNB}. Cross-tier resource disparity can be adjusted by defining the ratio of the transmit powers of two consecutive tiers, $\rho_i$, (i.e., $\rho_i = \frac{P_i}{P_{i+1}} > 1$), as well as the ratio of processing powers of their MEC hosts, $\gamma_i$, (i.e., $\gamma_i = \frac{C_i}{C_{i+1}} > 1$). It should be noted that the mentioned ratios are always greater than 1 as a tier $i \in \{1, \cdots, K\}$ is assumed to be overlaid with tiers of lower transmit power and processing capabilities. Note that $\rho_K = \gamma_K = 1$. 

\subsection{Signal and Channel Model}
The pathloss between a given \ac{UE} and its serving \ac{eNB} is modeled as inversely proportional to the distance with a given path-loss exponent denoted by $\alpha$, of common value for all tiers. Small-scale fading is assumed to be Rayleigh distributed with unit average power, i.e., for every \ac{UE}-\ac{eNB} link, $|h|^2 \approx \text{exp(1)}$ and the fast fading effects are assumed non-correlated among the various links. Additionally, each \ac{UE} employs a fixed transmit power, $P_{UE}$, which is greater than its serving \ac{eNB} sensitivity. The target \ac{eNB} belonging to the $i$-th tier is assumed to be placed at the origin \cite{Novlan2013}, thus, for \ac{UL} communication, the measured \ac{SINR} at that \ac{eNB} related to a transmission by the $k$-th \ac{UE} is 
\begin{equation}
\text{SINR}_{k,i} = \frac{P_{UE}|h_k|^2 D^{-\alpha}}{\hat{I}_z + \sigma^2}, 
\end{equation}
where $\sigma^2$ denotes the noise power, while $\hat{I}_z$ is the interference generated by an interfering set $\mathcal{Z}$ of \ac{UE}s as $\hat{I}_z = \sum_{z \in \mathcal{Z}} P_{UE} |h_z|^2 R_z ^{-\alpha}$. The random variables $D$ and $R_z$ represent the distance between the associated \ac{eNB} and the focused \ac{UE} and the distance between the same eNB and the interfering UEs, respectively. Finally, orthogonal channel allocation is assumed to avoid intra-cell interference. 

\subsection{Extended Packet Delay Budget}
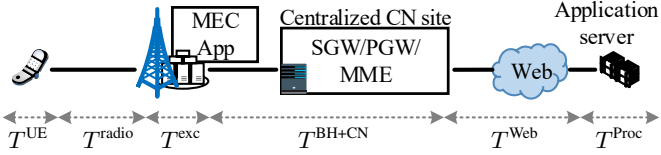
\begin{figure}
	\begin{center}
		\input{system_model/figures/system_latency/system_latency.tex}
		\caption{End-to-end latency overview.}
		\label{fig:system_latency}
	\end{center}
\end{figure}

As mentioned earlier, low latency access to cloud infrastructure is foreseen as a critical feature of 5G systems \cite{Ge2016}. As a result,  the experienced one-way \ac{E-PDB} at the \ac{UE} side during task offloading will be the focused metric throughout this work. As shown in Figure \ref{fig:system_latency}, the overall end-to-end \ac{E-PDB} is illustrated \cite{ETSISeptember2015}. First, $T^\text{UE}$ represents the time needed for application initiation and packet generation at the \ac{UE} side, followed by time intervals for data transmission and task execution at the \ac{MEC} host, denoted by $T^\text{radio}$ and $T^\text{exc}$, respectively. Throughout this work, $T^\text{UE}$ is implicitly modeled through $T^\text{radio}$, via random generation of packets, whereas, the back-haul, web and remote processing latencies, denoted by $T^\text{BH+CN}$, $T^\text{Web}$ and $T^\text{Proc}$ respectively, are assumed to be negligible. It is also assumed that the \ac{eNB}s and their corresponding \ac{MEC} hosts are physically located at the same node and that all deployed \ac{UE}s concurrently offload their tasks to their chosen \ac{MEC} host.

As modeled in \cite{Le2017}, the \ac{E-PDB} for the $k$-th \ac{UE} associated to an \ac{eNB} in the $i$-th tier is calculated as follows
\begin{equation}\label{eq:E-PDB_total}
\text{E-PDB}_{k,i} = T_{k,i}^{\text{radio}} + T_{k,i}^{\text{exc}},  
\end{equation}
where $T_{k,i}^{\text{radio}}$ and $T_{k,i}^{\text{exc}}$ stand for the radio propagation time and the task execution time at the \ac{MEC} host, respectively. The radio propagation latency represents the time needed for a given packet of size of $l_k$ bits to arrive at the serving \ac{eNB} \cite{Rajanna2017}, thus can be calculated as  
\begin{align}
T_{k,i}^{\text{radio}} &= \frac{l_k}{r_{k,i}} = \frac{l_k}{B_{k,i} \text{log}_2(1+\text{SINR}_{k,i})},
\end{align}
 where $r_{k,i}$ is the achievable rate of UE $k$ and $B_{k,i}$ represents the bandwidth allocated to UE $k$ when served by an \ac{eNB} in the $i$-th tier. On the other hand, the execution time can be computed as
 \begin{equation}
 T_{k,i}^{\text{exc}} = \frac{l_k f_k}{y_{k,i} C_i},
 \end{equation}
where $f_k$ (measured in cycles/bit) is the number of processing operations per input bit for the task to be offloaded by UE $k$ and $y_{k,i}$ represents the fraction of the total processing power of a tier-$i$ MEC host dedicated to the $k$-th \ac{UE}. 

Throughout this work, we assume equal per-user allocation of radio bandwidth and computational (MEC) resources \cite{Lin2015}, as the design of a more sophisticated resource allocation scheme is beyond the scope of this paper. Thus, for a given \ac{eNB} belonging to the $i$-th tier, the number of associated \ac{UE}s, which is obtained by means of applying a cell association rule, will determine the portion of bandwidth and computational resources dedicated to each connected \ac{UE}. In what follows, we present in detail the investigated association rules.  

%% file: system_model/figures/system_deployment/system_deployment_2.tex
\begingroup%
  \makeatletter%
  \providecommand\color[2][]{%
    \errmessage{(Inkscape) Color is used for the text in Inkscape, but the package 'color.sty' is not loaded}%
    \renewcommand\color[2][]{}%
  }%
  \providecommand\transparent[1]{%
    \errmessage{(Inkscape) Transparency is used (non-zero) for the text in Inkscape, but the package 'transparent.sty' is not loaded}%
    \renewcommand\transparent[1]{}%
  }%
  \providecommand\rotatebox[2]{#2}%
  \ifx\svgwidth\undefined%
    \setlength{\unitlength}{232bp}%
    \ifx\svgscale\undefined%
      \relax%
    \else%
      \setlength{\unitlength}{\unitlength * \real{\svgscale}}%
    \fi%
  \else%
    \setlength{\unitlength}{\svgwidth}%
  \fi%
  \global\let\svgwidth\undefined%
  \global\let\svgscale\undefined%
  \makeatother%
  \begin{picture}(1,0.78)%
    \put(0,0){\includegraphics[scale = 0.9]{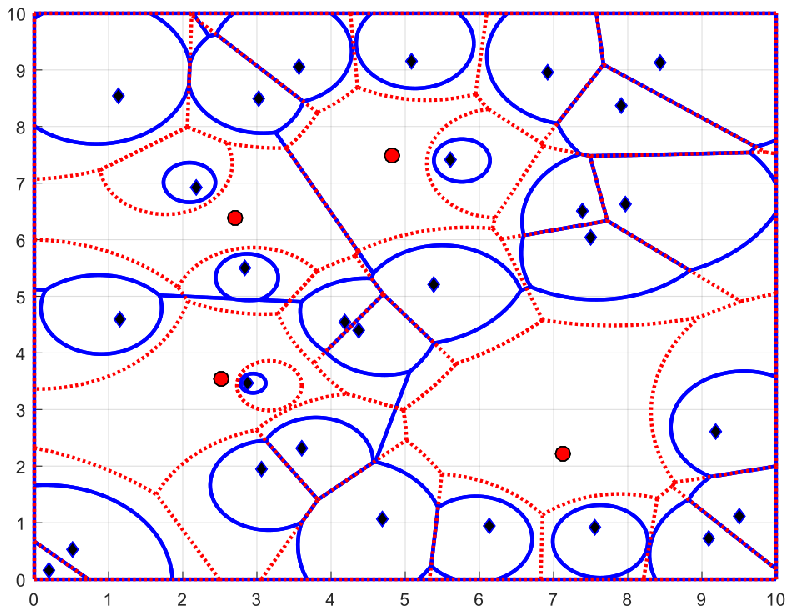}}%
    \put(0.02,0.27){\color[rgb]{0,0,0}\rotatebox{90.00000252}{\makebox(0,0)[lb]{\smash{y-coordinate [km]}}}}%
    \put(0.4,0){\color[rgb]{0,0,0}\makebox(0,0)[lb]{\smash{x-coordinate [km]}}}%
  \end{picture}%
\endgroup%

%% file: system_model/figures/system_latency/system_latency.tex
\begingroup%
  \makeatletter%
  \providecommand\color[2][]{%
    \errmessage{(Inkscape) Color is used for the text in Inkscape, but the package 'color.sty' is not loaded}%
    \renewcommand\color[2][]{}%
  }%
  \providecommand\transparent[1]{%
    \errmessage{(Inkscape) Transparency is used (non-zero) for the text in Inkscape, but the package 'transparent.sty' is not loaded}%
    \renewcommand\transparent[1]{}%
  }%
  \providecommand\rotatebox[2]{#2}%
  \ifx\svgwidth\undefined%
    \setlength{\unitlength}{280bp}%
    \ifx\svgscale\undefined%
      \relax%
    \else%
      \setlength{\unitlength}{\unitlength * \real{\svgscale}}%
    \fi%
  \else%
    \setlength{\unitlength}{\svgwidth}%
  \fi%
  \global\let\svgwidth\undefined%
  \global\let\svgscale\undefined%
  \makeatother%
  \begin{picture}(1,0.195)%
    \put(0,0){\includegraphics[width=\columnwidth]{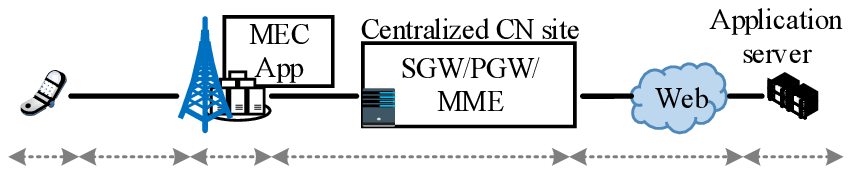}}%
    \put(0.01,-0.03){\color[rgb]{0,0,0}\makebox(0,0)[lb]{\small{$T^\text{UE}$}}}%
    \put(0.11,-0.03){\color[rgb]{0,0,0}\makebox(0,0)[lb]{\small{$T^\text{radio}$}}}%
    \put(0.215,-0.03){\color[rgb]{0,0,0}\makebox(0,0)[lb]{\small{$T^\text{exc}$}}}%
    \put(0.4,-0.03){\color[rgb]{0,0,0}\makebox(0,0)[lb]{\small{$T^\text{BH+CN}$}}}%
    \put(0.66,-0.03){\color[rgb]{0,0,0}\makebox(0,0)[lb]{\small{$T^\text{Web}$}}}%
    \put(0.8,-0.03){\color[rgb]{0,0,0}\makebox(0,0)[lb]{\small{$T^\text{Proc}$}}}%
  \end{picture}%
\endgroup%

%% file: flexible_cell_association/flexible_cell_association.tex
\section{Flexible Cell Association}\label{sec:flexible_cell_association}
Radio-based achievable gains, with reference to \ac{UL} rates, load balancing and system throughput have been shown in \cite{Boccardi2016}, \cite{Singh2015}, however, the \ac{UL} cell association is achieved based on an \ac{eNB} proximity criterion, hence, leading to the minimum pathloss experienced by the \ac{UE}. In this work, we choose to revisit this rule and propose a new, \ac{MEC}-aware cell association rule, that aims at minimizing the execution time at the \ac{MEC} host, along with ensuring connectivity to the closest \ac{eNB}. This is motivated through questioning the optimality of the conventional, maximum \ac{DL} \ac{RSRP}-based association rule,  when it comes to the task offloading latency experienced by a \ac{UE} in a \ac{HetNet}. A mathematical representation of the association problem can be formulated as follows
\begin{align}\label{eq:association_rule}
x_i =& \argmax_{x \in \bm{\Phi_i}} (\eta_i || x-y ||^{-\alpha}), \forall i = 1, 2, \cdots, K,\\ \nonumber
x_o =& \argmax_{x \in x_i: i = 1,\cdots K}  (\eta_i || x-y ||^{-\alpha}),
\end{align}
where $\eta_i, i=1, \cdots, K$ represents a biasing factor for the $i$-th tier imposed to the \ac{UE}s, $y$ is the \ac{UE}'s location and the $||.||$ operation denotes the Euclidean distance between two points of the two-dimensional plane. With regards to the choice of values for parameter $\eta_i$, the conventional and proposed association rules are discussed in the remainder of this section. 

\subsection{Reference Signal Received Power Criterion}
According to this cell association rule, a \ac{UE} is served by the \ac{eNB} providing it with the maximum \ac{RSRP} in the \ac{DL}. This is equivalent to setting $\eta_i$ to be equal to $P_i$ in Eq. (\ref{eq:association_rule}). 

In a highly heterogeneous \ac{HetNet} of large radio disparity (i.e., $\rho_i >> 1$), execution of this rule leads to an imbalanced load among the multiple tiers and, as a result, to limited radio performance, since most of the \ac{UE}s will be associated to \ac{eNB}s of high total transmit power. This problem is well-known and multiple solutions have been proposed, such as load-aware optimization \cite{Ge2016} and cell range extension \cite{Holma2012}. In order to quantify the number of \ac{UE}s associated with a tier-$i$ \ac{eNB}, the association probability of a given \ac{UE} to an \ac{eNB} of the $i$-th tier is calculated as  $\mathcal{A}_i^{\text{RSRP}} = \frac{\lambda_i}{\Delta^{\text{RSRP}}_i}$, where $\Delta^{\text{RSRP}}_i = P_i^{\frac{-2}{\alpha}} \sum_{j=1}^{K}\lambda_j P_j^{\frac{2}{\alpha}}$ \cite{SakrHossain2014}. Consequently, the average number of associated \ac{UE}s to an \ac{eNB} of the $i$-th tier, termed as $\hat{N}^{\text{RSRP}}_i$, will affect the experienced \ac{E-PDB} per UE, as the amount of bandwidth and processing resources allocated per UE is inversely proportional to the achieved \ac{E-PDB}. Mathematically, quantity $\hat{N}^{\text{RSRP}}_i$ is expressed as 
\begin{equation}
\hat{N}^{\text{RSRP}}_i =  \frac{\mathcal{A}_i^{\text{RSRP}} \lambda_u}{\lambda_i}  = \frac{\lambda_u}{\Delta^{\text{RSRP}}_i}.
\end{equation}
Assuming equal resource allocation among the \ac{UE}s connected to an \ac{eNB}, the bandwidth and processing (MEC) resources allocated to the $k$-th \ac{UE} associated to an \ac{eNB} of the $i$-th tier will be equal to
\begin{align}
	B_{k,i} =& \frac{B_{\text{i}}}{\hat{N}^{\text{RSRP}}_i}, \\
	y_{k,i}  =& \frac{1}{\hat{N}^\text{RSRP}_i}, 
\end{align}
where $B_{\text{i}}$ represents the total bandwidth allocated to tier $i, i=1, \cdots, K$.

\subsection{Computational Proximity Criterion}
In what follows, we introduce a new, \ac{MEC}-aware cell association rule, according to which the serving \ac{eNB} is the one of the maximum \emph{computational proximity}. In this context, computational proximity refers to the existence of a processing power source in the vicinity of a device of limited computation capabilities that chooses to offload a demanding task to this source. Such resources, as defined in Section \ref{sec:system_model} ($C_i, \forall i=1,\cdots,K$) can be the same for all the tiers, thus, resulting in a homogeneous network from a \ac{MEC} perspective, or can be varying across the tiers, resulting in a MEC \ac{HetNet}, thus, affecting the task offloading latency experienced by a \ac{UE}. As observed from the total \ac{E-PDB} expression in Eq. (\ref{eq:E-PDB_total}), the overall \ac{E-PDB} is jointly affected by the proximity to the connected \ac{eNB} (i.e., radio part - $T_{k,i}^{\text{radio}}$) as well as by the available processing power (i.e., \ac{MEC} part - $T_{k,i}^{\text{exc}}$). Our aim is to consider both resource domains through introducing a new association rule for \ac{UL} communication, by setting the bias factors $\eta_i$ as functions of the available computational resources (i.e., $\eta_i = C_i$). As a consequence, the association probabilities and the average numbers of connected users can be computed easily by replacing $P_i$ by $C_i$ and computing $\mathcal{A}_i^{\text{MEC}}$, $\Delta^{\text{MEC}}_i$ and $\hat{N}^{\text{MEC}}_i$, accordingly. 
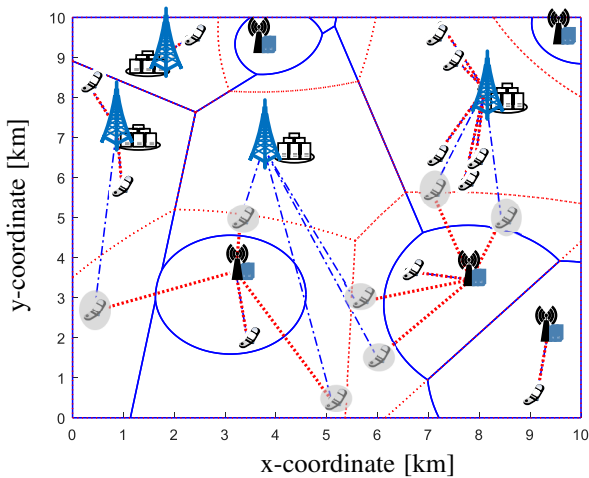
\begin{figure}[t]
	\begin{center}
		\input{flexible_cell_association/figures/coverage_regions/coverage_regions.tex}
		\caption{A zoomed realization of a two-tier network consisting of macro and micro eNBs ($\rho_1 = 40,\; \gamma_1 = 2, \; \omega_1 = 20$). The blue dashed-dotted lines represent \ac{UE} connectivity following the maximum \ac{DL} \ac{RSRP} association rule, while, the red dashed lines represent \ac{UE} connectivity based on the proposed computational proximity-based rule.The gray shaded users are the ones for which execution of the two cell association rules results to different \ac{eNB}/ \ac{MEC} nodes for connectivity.}
		\label{fig:coverage_regions}
	\end{center}
\end{figure}

Referring to network deployment, as it will be shown, a critical factor affecting the performance of the proposed \ac{UE}-cell association rule is the ratio of radio/ MEC cross-tier disparities, which is defined as
\begin{equation}\label{eq:omega}
\omega_i = \frac{\rho_i}{\gamma_i}.
\end{equation}
In order to visualize the influence of parameter $\omega_i$ on \ac{UE} connectivity, focusing on a two-tier network, Figure \ref{fig:coverage_regions} presents a zoomed overview of a network realization, where \ac{UE}s are connected to their serving \ac{eNB}s/\ac{MEC} hosts via the two discussed rules. One can observe that, assuming a large value of parameter $\omega_1$, for a fair number of \ac{UE}s, the maximum \ac{DL} \ac{RSRP} association rule indicates a node for connectivity which is different from the one obtained by applying the proposed computational proximity-based association rule. This occurs because large cross-tier radio/ MEC disparities lead towards quite dissimilar radio/ MEC coverage areas. Such an observation paves the way towards a different insight on the network planning process, taking into account the available computational resources together with the radio transmission capabilities, since both of them directly affect the \ac{E-PDB} experienced by a given \ac{UE}, when the latter wishes to offload a demanding processing task to a \ac{MEC} host.

In the following section, we present various simulation results, highlighting key messages regarding the studied association rules, the role of cross-tier parameter disparities, as well as the effect of deployment densities on the achieved \ac{E-PDB}.

%% file: flexible_cell_association/figures/coverage_regions/coverage_regions.tex
\begingroup%
  \makeatletter%
  \providecommand\color[2][]{%
    \errmessage{(Inkscape) Color is used for the text in Inkscape, but the package 'color.sty' is not loaded}%
    \renewcommand\color[2][]{}%
  }%
  \providecommand\transparent[1]{%
    \errmessage{(Inkscape) Transparency is used (non-zero) for the text in Inkscape, but the package 'transparent.sty' is not loaded}%
    \renewcommand\transparent[1]{}%
  }%
  \providecommand\rotatebox[2]{#2}%
  \ifx\svgwidth\undefined%
    \setlength{\unitlength}{222bp}%
    \ifx\svgscale\undefined%
      \relax%
    \else%
      \setlength{\unitlength}{\unitlength * \real{\svgscale}}%
    \fi%
  \else%
    \setlength{\unitlength}{\svgwidth}%
  \fi%
  \global\let\svgwidth\undefined%
  \global\let\svgscale\undefined%
  \makeatother%
  \begin{picture}(1,0.81115108)%
    \put(0,0){\includegraphics[scale= 0.9]{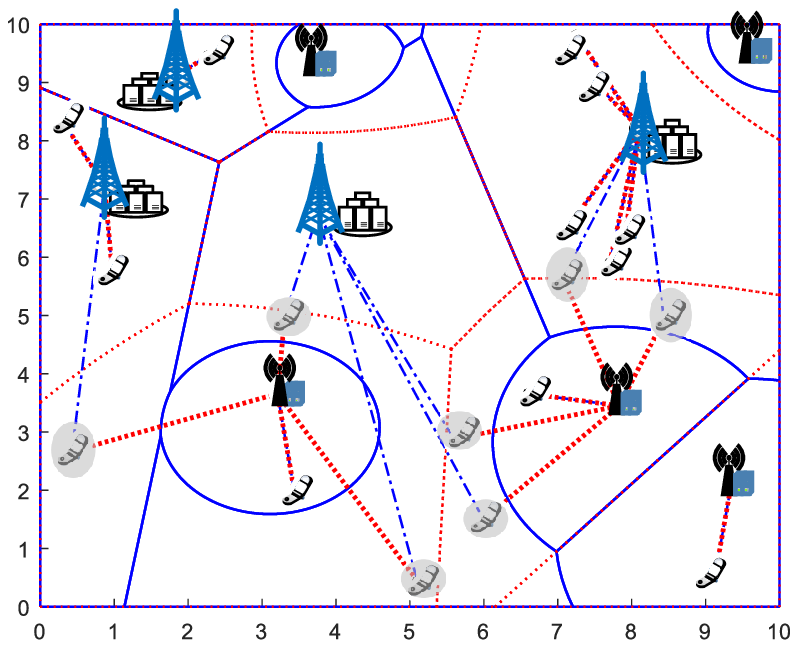}}%
    \put(0,0.27){\color[rgb]{0,0,0}\rotatebox{90.00000252}{\makebox(0,0)[lb]{\smash{y-coordinate [km]}}}}%
    \put(0.4,0){\color[rgb]{0,0,0}\makebox(0,0)[lb]{\smash{x-coordinate [km]}}}%
  \end{picture}%
\endgroup%

%% file: simulation_results/simulation_results.tex
\section{Simulation Results}\label{sec:simulation_results}

Our objective throughout this section is to provide insight on the \ac{E-PDB} improvements when applying the new proposed \ac{MEC}-aware association rule, by means of numerical evaluation, as an analytical \ac{E-PDB} study is planned for future work. A two-tier \ac{HetNet} is investigated, where the $k$-th \ac{UE} generates a random packet of size of $l_k$ bits that is modeled as a uniform random variable taking values between $l_\text{min}$ and $l_\text{max}$. Additionally, the number of processing operations per input bit, $f_k$, is uniformly distributed, as well, between values $f_\text{min}$ and $f_\text{max}$. The amount of dedicated bandwidth and computational resources that each \ac{eNB} assigns to its associated users is computed based on the applied association rules. Due to the utilized random spatial model (i.e., \ac{PPP}-based), Monte-Carlo simulations were conducted for the \ac{eNB}/ \ac{UE} locations and the small-scale fading phenomena. A summary of the adopted simulation parameters is provided in Table \ref{Table:simulation_parameters}, where the parameter values are fixed throughout the section, unless otherwise stated. It should be noted that, as a two-tier \ac{HetNet} is considered, the subscript of parameter $\omega_1$ will be dropped for the sake of simplicity.  
\begin{table}[!t]
	\centering
	\caption{Simulation Parameters}
	\renewcommand{\arraystretch}{1}
	\resizebox{0.9\columnwidth}{!}{
		\begin{tabular}{| l | l |}
			\hline
			Parameter & value  \\ \hline \hline
			Number of tiers & 2 \\
			$(\lambda_1, \lambda_2)$ & (0.5, 3) \ac{eNB}s/km \\
			$(P_1, P_2)$ & (46, 30) dBm \\
			Relative area & 10 $\text{km}^2$ \\ 
			$\lambda_u$ & 30 \ac{UE}s/km \\
			$P_{\text{UE}}$ & 23 dBm \\ 
			$\sigma^2$ & -90 dBm  \\  
			$(l_\text{min}, l_\text{max})$& (100, 300) kbits \\ 
			$(f_\text{min}, f_\text{max})$& (500, 1500) cycles/bit \\
		 	Bandwidth & 10 MHz\\
		 	Pathloss exponent ($\alpha$) & 4 \\
			Number of realizations & 10000 \\ \hline
	\end{tabular}}
	\label{Table:simulation_parameters}
\end{table}

\begin{figure}
	\begin{center}
		\input{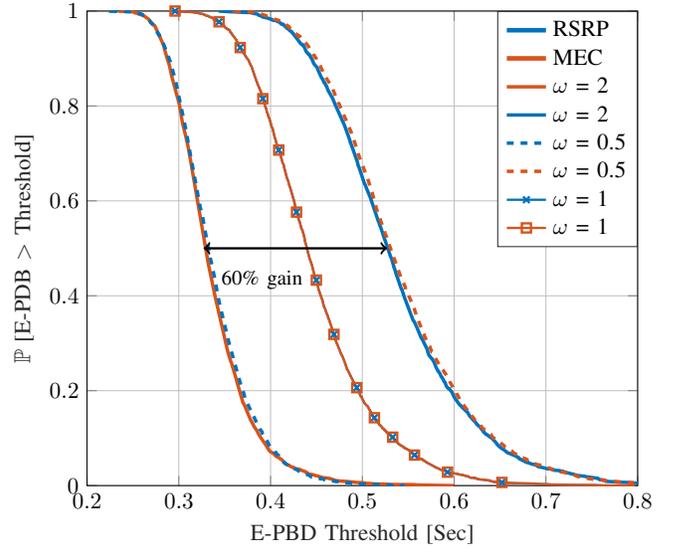}
		\vspace{-15pt}
		\caption{\ac{CCDF} of the \ac{E-PDB} under different radio and \ac{MEC} resource disparities.}
		\label{fig:E-PDB_CCDF}
	\end{center}
\end{figure}

As mentioned in Section \ref{sec:flexible_cell_association}, the achievable \ac{E-PDB} is our metric of investigation throughout this work. In Figure \ref{fig:E-PDB_CCDF}, the \ac{CCDF} of the \ac{E-PDB} is shown for the two discussed association rules and for different values of $\omega$. As previously explained, when $\omega$ varies away from the value of one, the radio and \ac{MEC} coverage areas become more dissimilar, hence, resulting in a selection divergence of the associating \ac{eNB}/\ac{MEC} server by a \ac{UE}. It is observed that, for values of $\omega$ greater than one ($\omega$ = 2), the proposed computational proximity association rule (denoted by ``\ac{MEC}'') provides a lower probability to violate a given \ac{E-PDB} threshold as compared to the maximum \ac{RSRP} rule (denoted by ``\ac{RSRP}''), with nearly 60\% \ac{E-PDB} reduction for the 50-th percentile of UEs. This occurs due to the enhanced balance between the proximity and available computational resources at the \ac{MEC} node. On the other hand, as $\omega$ is lower than one ($\omega$ = 0.5), the performance is turned over, as the \ac{RSRP} rule provides a lower experienced \ac{E-PDB} of the same latency reduction. Consequently, we observe that having the two association metrics at hand, an adaptive, deployment-dependent cell association procedure can be envisioned, in order to fully capture the radio and \ac{MEC} resource disparities for \ac{E-PDB} minimization. Under that framework, the \ac{UE} is ought to only acquire knowledge of the  radio and \ac{MEC} disparities of the \ac{HetNet}, in order to decide upon which association rule to consider. For the case of $\omega = 1$, since the corresponding coverage areas obtained by the two rules will fully overlap, the experienced \ac{E-PDB} performance will be identical for the two rules.   

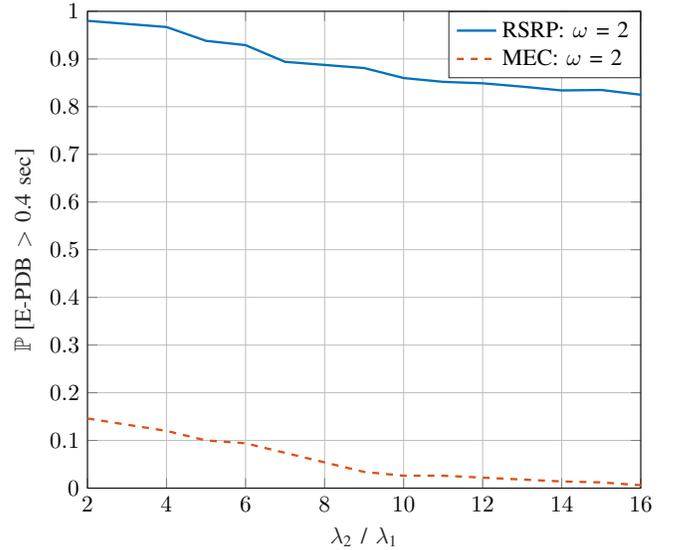
\begin{figure}
	\begin{center}
		\input{simulation_results/figures/PDB_spatial_density/PDB_spatial_density.tex}
		\vspace{-15pt}
		\caption{ Probability to violate a target \ac{E-PDB} (0.4 sec) as a function of the ratio of \ac{eNB}/\ac{MEC} deployment densities.}
		\label{fig:E-PDB_spatial_density}
	\end{center}
\end{figure}

With the aim of observing the effect of deployment density on the experienced \ac{E-PDB}, Figure \ref{fig:E-PDB_spatial_density} depicts the probability of violating a target \ac{E-PDB} of 0.4 seconds for an increasing ratio of micro-over-macro \ac{eNB} spatial densities when $\omega = 2$. We observe a nearly constant association-based outage reduction in favor of the proposed \ac{MEC}-aware association rule, similar to the latency reduction observed in Figure \ref{fig:E-PDB_CCDF}. The decreasing slope of the two curves is expected as the number of micro \ac{eNB}s over a unit area increases. This is due to the increasing probability for a \ac{UE} to be associated with a closer node, thus leading to lower \ac{E-PDB} values. 

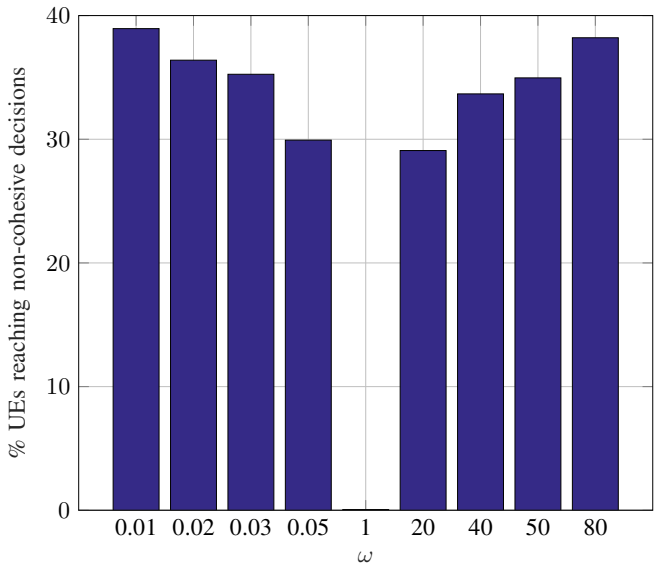
\begin{figure}
	\begin{center}
		\input{simulation_results/figures/percentile_decoupled/Percentile_Decoupled.tex}
		\vspace{-15pt}
		\caption{Fraction of \ac{UE}s reaching non-cohesive decisions upon cell association, as a function of cross-tier radio and \ac{MEC} disparity.}
		\label{fig:percentile_decoupled}
	\end{center}
\end{figure}

Finally, in Figure \ref{fig:percentile_decoupled}, the percentage of \ac{UE}s for which the maximum \ac{DL} \ac{RSRP} and the proposed \ac{MEC}-aware cell association rules provide different connectivity recommendations, is illustrated, as a function of the value of parameter $\omega$. As anticipated, for the increase of cross-tier disparity between the radio and \ac{MEC} capabilities (i.e. $\omega \neq 1$), the two coverage areas become highly divergent, thus, leading to a higher probability of a \ac{UE} being present in this disjoint region (e.g., nearly 40 \% of UEs will reach different decisions upon associating to an \ac{eNB}/ \ac{MEC} node for large disparities of $\omega= 0.01$ or $\omega= 80$). On the contrary, for the $\omega = 1$ case, the radio and \ac{MEC} coverage areas will be identical, hence, the application of the two investigated association rules will provide the same preference for \ac{UL} connectivity. 
	
%% file: simulation_results/figures/PDB_spatial_density/PDB_spatial_density.tex
%
%
\definecolor{mycolor1}{rgb}{0.00000,0.44700,0.74100}%
\definecolor{mycolor2}{rgb}{0.85000,0.32500,0.09800}%
 \resizebox {\columnwidth} {!} {
\begin{tikzpicture}

\begin{axis}[%
width==0.94*8cm,
height==0.94*8cm,
at={(0.758in,0.519in)},
scale only axis,
xmin=2,
xmax=16,
xlabel style={font=\color{white!15!black}},
xlabel={$\lambda{}_\text{2}\text{ / }\lambda{}_\text{1}$},
ymin=0,
ymax=1,
ytick = {0, 0.1, 0.2, 0.3, 0.4, 0.5, 0.6, 0.7, 0.8, 0.9, 1},
ylabel style={font=\color{white!15!black}},
ylabel={$\mathbb{P}\;\text{[E-PDB \textgreater{} 0.4 sec]}$},
axis background/.style={fill=white},
xmajorgrids,
ymajorgrids,
legend style={at={(1,1)},anchor=north east, legend cell align=left, align=left, draw=white!15!black}
]
\addplot [color=mycolor1, line width=1.0pt,]
  table[row sep=crcr]{%
2	0.98\\
4	0.967\\
5	0.938\\
6	0.929\\
7	0.894\\
9	0.881\\
10	0.86 \\
11	0.852\\
12	0.849\\
13	0.842\\
14	0.834\\
15  0.835\\
16  0.825\\
17  0.827\\
18  0.818\\
};
\addlegendentry{$\text{RSRP: } \omega\text{ = 2}$}

\addplot [color=mycolor2, dashed, line width=1.0pt]
  table[row sep=crcr]{%
2	0.146\\
4	0.12\\
5	0.1\\
6	0.094\\
9	0.034\\
10	0.026\\
11	0.026\\
14	0.014\\
15	0.012\\
16  0.006\\
17  0.006\\
18  0.003\\
};
\addlegendentry{$\text{MEC: } \omega\text{ = 2}$}
\end{axis}
\end{tikzpicture}%
}

%% file: simulation_results/figures/percentile_decoupled/Percentile_Decoupled.tex
%
%
\definecolor{mycolor1}{rgb}{0.20810,0.16630,0.52920}%
\resizebox {\columnwidth} {!} {
	\begin{tikzpicture}
	
	\begin{axis}[%
	width==0.94*8cm,
	height==0.94*8cm,
	at={(0.758in,0.519in)},
	scale only axis,
	bar shift auto,
	xmin=0,
	xmax=10,
	xtick={1,2,3,4,5,6,7,8,9},
	xticklabels={{0.01},{0.02},{0.03},{0.05},{1},{20},{40},{50},{80}},
	xlabel style={font=\color{white!15!black}},
	xlabel={$\omega$},
	ymin=0,
	ymax=40,
	ylabel style={font=\color{white!15!black}},
	ylabel={\% UEs reaching non-cohesive decisions},
	axis background/.style={fill=white},
	xmajorgrids,
	ymajorgrids
	]
	\addplot[ybar, bar width=0.8, fill=mycolor1, draw=black, area legend] table[row sep=crcr] {%
		1	38.9446161539388\\
		2	36.3934907473253\\
		3	35.2546754095822\\
		4	29.9290486791057\\
		5	0.0573942317902807\\
		6	29.0877151576703\\
		7	33.6675491598125\\
		8	34.9617168207841\\
		9	38.2061363378007\\
	};
	\addplot[forget plot, color=white!15!black] table[row sep=crcr] {%
		0	0\\
		10	0\\
	};
	\end{axis}
	\end{tikzpicture}%
}

%% file: conclusion/conclusion.tex
\section{Conclusion}\label{sec:Conclusion}
In this work we have leveraged the \ac{MEC} degree of freedom in planning and dimensioning a \ac{HetNet}, via optimizing the exploitation of both communication and computation resources during \ac{UE}-cell association. Focusing on the task offloading example, a new association metric for \ac{UL} communication has been proposed, aiming at reducing the experienced \ac{E-PDB} of a \ac{UE}. Different scenarios spanning diverse radio and \ac{MEC} cross-tier disparities have been presented to highlight the cell association decision effect on system performance. It has been shown that, for a range of disparities between radio and \ac{MEC} capabilities between tiers, the proposed computational proximity rule provided gains in terms of \ac{E-PDB}, as compared to the conventional maximum \ac{RSRP} rule. This performance gain degrades as cross-tier radio/ \ac{MEC} disparities become similar. Also importantly, we have explored the case, in which, for different association rules, a \ac{UE} would favor associating to different \ac{eNB}/\ac{MEC} hosts in the \ac{UL}. 

\section*{Acknowledgment}
The research leading to these results has been performed under the framework of the Horizon 2020 project ONE5G (ICT-760809) receiving funds from the European Union. The authors  would like to acknowledge the contributions of the colleagues in the project, although the views and work expressed in this contribution are those of the authors and do not necessarily represent the project.

%% file: literature/literature.tex
\bibliographystyle{./lib/IEEEtran.cls}
\bibliography{./literature/Literature_Local}